# Metallurgical synthesis methods for Mg-Al-Ca scientific model materials

W. Luo[a], L. Tanure[a], M. Felten[b], J. Nowak[b], W. Delis[c], M. Freund[c], N. Ayeb[a], D. Zander[b], C. Thomas[d], M. Feuerbacher[d], S. Sandlöbes-Haut[c], S. Korte-Kerzel[c], H. Springer[a,e,*]

[a] RWTH Aachen University, Institute of Metal Forming, Intzestrasse 10, 52072 Aachen, Germany
[b] RWTH Aachen University, Corrosion and Corrosion Protection, Intzestrasse 5, 52072 Aachen, Germany
[c] RWTH Aachen University, Institute of Physical Metallurgy and Metal Physics, Kopernikusstrasse 14, 52074 Aachen, Germany
[d] Forschungszentrum Jülich GmbH, ER-C-1, Wilhelm-Johnen-Straße, 52428 Jülich, Germany
[e] Max-Planck-Institut für Eisenforschung GmbH, Max-Planck-Strasse 1, 40237 Düsseldorf, Germany

[*] Corresponding author
E-mail address: hauke.springer@ibf.rwth-aachen.de

**Abstract**

Mg-based alloys are industrially used for structural applications, both as solid solutions alloys and as composites containing intermetallic compounds. However, a further development in terms of mechanical properties requires the investigation of underlying causalities between synthesis, processing and microstructure to adjust the mechanical and the corrosion properties, ideally down to the near atomic scale. Such fundamental scientific investigations with high resolution characterisation techniques require model materials of exceptionally high purity and strictly controlled microstructure e.g. with respect to grain size, morphology, chemical homogeneity as well as content and size of oxide inclusions. In this context, the Mg-Al-Ca system appears exceptionally challenging from a metallurgical perspective due to the high reactivity and high vapor pressures, so that conventional industrial techniques cannot be successfully deployed. Here, we demonstrate the applicability of various scientific synthesis methods from arc melting over solution growth to diffusion couples, extending to effects and parameters for thermo-mechanical processing. Suitable pathways to overcome the specific challenges of the Mg-Al-Ca system are demonstrated, as well as the persistent limitations of the current state of the art laboratory metallurgy technology.

# 1 Introduction

The application of light-weight components in the aeronautics and automotive industry has both environmental and economic benefits [1]. Mg-based alloys are especially attractive due to their low density compared to steels, aluminium, and even some polymer-based materials. However, they typically suffer from a limited strength and especially low ductility [2]. Recent studies showed that Mg-based solid solution alloys with a small amount of Ca and Al (e.g. Mg-1Al-0.1Ca (wt.%)) exhibit improved formability and ductility due to the activation of $\langle c+a \rangle$ slip [3]. With increasing Al and Ca content, different types of intermetallic phases precipitate in the Mg-Al-Ca alloys and composites consisting of a Mg matrix and an intermetallic skeleton are obtained [4]. The precipitate changes from $Mg_{17}Al_{12}$ to the C15-$CaAl_2$, C36-$Ca(Mg,Al)_2$ and C14-$CaMg_2$ Laves phases with increasing Ca/Al ratio [5,6]. The continuous Laves phase network can reinforce the soft Mg matrix and improve the creep resistance of the Mg-Al-Ca alloys [7-9]. However, the knowledge of the mechanical properties [10-12] and corrosion properties [13] of the Laves phase is still limited. A systematic investigation of the Mg-Al-Ca solid solutions, composites and intermetallic phases will therefore facilitate the development of advanced Mg alloys for structural applications.

The main challenges in the production and processing of Mg-Al-Ca alloys of solid solutions and composites are related to the high reactivity [14] and high vapor pressures [15,16] of Mg and Ca. Although the synthesis of Mg-Al-Ca alloys based on industrial standards is feasible [17,18], it is more challenging to prepare the Mg-Al-Ca alloys of solid solutions and composites for specific scientific purposes, i.e. achieving very high purity, precisely controlling the phases, grain sizes and microstructures. Induction melting under protective atmosphere with high pressure can reduce evaporation of Mg and Ca during melting. Since Mg and Ca do not dissolve Fe during melting, Mg-Ca alloys can be melted and held in a crucible made of steel. However, as Al has a very high affinity with Fe [19], the molten Mg-Al-Ca alloys may react with the ferrous crucible, leading to Fe contamination in the Mg-Al-Ca alloys. Furthermore, phenomena such as grain boundary precipitation and intergranular segregation in the as-cast Mg alloys need to be reduced by further thermal mechanical processing [20].

In addition to the solid solutions and composites of the Mg-Al-Ca system, we also consider the binary [21-23] and ternary [24] intermetallic phases of the Mg-Al-Ca system. While induction melting using a steel crucible under Ar atmosphere is applicable to the synthesis of C14-$CaMg_2$ Laves phase, it is not suitable for the synthesis of the C15-$CaAl_2$ Laves phase due to Fe contamination. Alternatively, as there is a lack of Mg, we might be able to prepare the C15-$CaAl_2$ Laves phase by arc melting without causing serious evaporation. Due to the distinct brittleness of the Laves phase [25], the alloys of the C14-$CaMg_2$ and C15-$CaAl_2$ Laves phases might disintegrate during cooling. Casting into a pre-heated alumina crucible followed by slow cooling [26] was reported to reduce thermal stress during cooling. However, alumina crucibles cannot be used in the synthesis of the C14-$CaMg_2$ and C15-$CaAl_2$ Laves phases due to the high reactivity of Ca. It has been reported that $Mg_{17}Al_{12}$ intermetallic phase with a small amount of a second phase can be produced by using an electrical resistance melting furnace in a graphite crucible under protective Ar atmosphere [27]. Although homogenization heat treatment reduced the amount of the second phase, it leads to the formation of porosity in the alloys [27]. In order to obtain pure $Mg_{17}Al_{12}$ phase, we prepared a Mg-Al diffusion couple in the present work. It is even more challenging to synthesize the C36-$Ca(Mg,Al)_2$ Laves phase which has high Mg and Ca contents and narrow composition and temperature ranges [24,28]. The C36-$Ca(Mg,Al)_2$ Laves phase was only observed in Mg-Al-Ca alloys containing multiple phases [24,28]. To synthesize single-phase C36-$Ca(Mg,Al)_2$ Laves phase, we need to overcome not only the high vapor pressures and reactivity of Mg and Ca but also the difficulties in control of composition and temperature. Different methods such as the manual induction melting, diffusion couple technique, flux-growth method, and Bridgman method were utilized to synthesize the C36-$Ca(Mg,Al)_2$ Laves phase. In summary, the aim of the present work is to explore the challenges and pitfalls in the synthesis of the Mg-Al-Ca bulk materials of solid solution, composite and intermetallic phases, and to discuss solutions to the synthesis. An overview of metallographic preparation methods for the Mg-Al-Ca alloys has been published elsewhere [29]. We hope that these methods of synthesis and metallographic preparation could help other researchers working on the Mg-Al-Ca alloys.

## 2 Materials and methods

The Mg-Al-Ca alloys were prepared from pure Mg (99.95 wt.%), Ca (98.8 wt.%) and Al (99.999 wt.%). To overcome the challenges in synthesis, different methods including induction melting, arc-melting, diffusion couple technique, manual induction melting, flux-growth method and Bridgman method, were used to produce the Mg-Al-Ca alloys of solid solutions, composites and intermetallic phases (Table 1). The chemical analyses of the starting materials and the synthesized alloys were performed by inductively coupled plasma optical emission spectroscopy (ICP-OES). The metallographic preparation methods for the Mg-Al-Ca alloys are explained in details elsewhere [29]. The microstructures of the alloys were observed using scanning electron microscopy (SEM) (Helios Nanolab 600i, FEI Inc.). The compositions of the phases in the alloys were determined by energy dispersive X-ray spectroscopy (EDS) (EDAX

Inc.). Phase analyses were carried out by electron backscatter diffraction (EBSD) (Hikari, EDAX Inc.). Some samples were colour etched with a picric acid-based solution for optical microscopy (OM) (Leica DMR, Leica AG) observations.

*Table 1 Summary of the metallurgical synthesis methods for the Mg-Al-Ca solid solutions, composites and intermetallics.*

| Materials | Composition (wt.%) | Phase | Method | Heat treatment/ thermal mechanical processing |
|---|---|---|---|---|
| Mg-Al-Ca solid solution | Mg-1Al-0.5Ca<br>Mg-1Al-0.05Ca<br>Mg-1Al-0.1Ca<br>Mg-1Al-0.2Ca<br>Mg-2Al-0.05Ca<br>Mg-2Al-0.1Ca<br>Mg-2Al-0.2Ca | Mg | Induction melting under Ar atmosphere at 10 bar using a steel crucible and casting under Ar atmosphere at 15 bar in a Cu mould | Hot-rolling at 450 °C in five passes with 10 % reduction in thickness per pass followed by annealing at 500 °C for 24 h |
| Mg-Al-Ca composite | Mg-6Al-2Ca<br>Mg-5Al-3Ca<br>Mg-4Al-4Ca | Mg + Laves phase | Induction melting under Ar atmosphere at 0.8 bar using a steel crucible and casting under Ar atmosphere at 0.8 bar in a Cu mould | Annealed under Ar atmosphere at 500 °C for 48 h, followed by furnace cooling |
| Mg-Al-Ca intermetallic | Mg-45Ca | C14-CaMg$_2$ | Induction melting under Ar atmosphere at 0.8 bar using a steel crucible and casting under Ar atmosphere at 0.8 bar in a Cu mould | As-cast |
| | Al-42Ca | C15-CaAl$_2$ | Arc-melting on a water-cooled copper hearth under Ar atmosphere | Annealed in a glass tube furnace at 600 °C for 24 h under Ar atmosphere |
| | Mg-28Al-46Ca | C36-Ca(Mg,Al)$_2$ | Pure elements were placed in a cylindrical tantalum crucible sealed with 0.6 bar Ar, placed on a water-cooled cold finger in a Bridgman apparatus and then heated to 900 °C. The growth was carried out by lowering the crucible out of the hot zone at a velocity of 5 mm/h. | As-cast |

# 3 Results and discussion

## 3.1 Mg-Al-Ca solid solutions

In the present work, the Mg-Al-Ca solid solutions were melted in a 60 kW induction furnace using a steel crucibles under Ar atmosphere with pressure of 10 bar. As the Al content is only 1-2 wt.% in the Mg-Al-Ca solid solutions, solid solution of Fe in the melt is considered to be negligible (0.0068 wt.% Fe measured by ICP-OES). During melting, the inductive field couples with the ferromagnetic crucible and stirs the melt in the steel crucible, which facilitates homogenization of the melt. In order to reduce the casting defects, the Mg-Al-Ca solid solutions were cast under Ar atmosphere with pressure of 15 bar. The melt was cast into a copper mould with an internal cross section 30 × 150 mm$^2$ and an ingot of about 500 g was obtained. As shown in Fig. 1(a), the Mg-1Al-0.05Ca (wt.%) alloy cast under high pressure Ar exhibit only a small amount of pores. Tensile tests (Fig. 1(b)) further reveal that the Mg-1Al-0.05Ca (wt.%) alloy cast under Ar atmosphere with pressure of 15 bar exhibit higher tensile strength and ductility than those cast under Ar atmosphere with pressure of 0.8 bar, as a result of the reduced defect density.

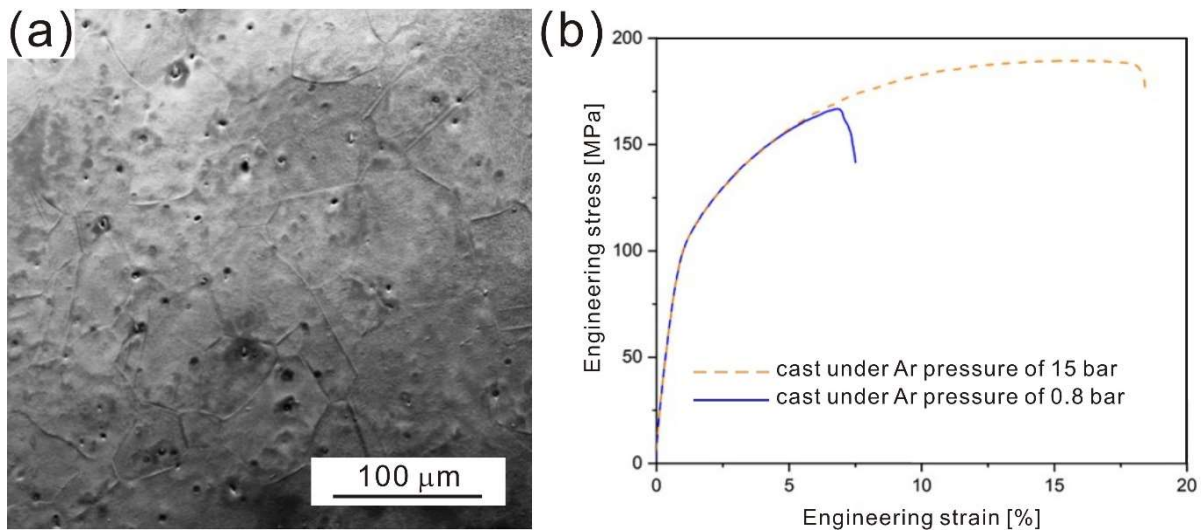

*Fig. 1 (a) SEM image of the Mg-1Al-0.05Ca (wt.%) alloy cast under Ar atmosphere with pressure of 15 bar. (b) Engineering stress-strain curves of the tensile tests of the Mg-1Al-0.05Ca (wt.%) alloys cast under Ar atmosphere with pressure of 15 bar and 0.8 bar.*

Furthermore, the as-cast Mg-Al-Ca solid-solution alloys were subjected to thermal mechanical processing by hot rolling followed by homogenization and recrystallization annealing to reduce the shrinkage cavities and porosity caused during solidification, and to homogenise the elements distribution. The temperatures of the thermal mechanical processing including the annealing temperature and rolling temperature have been selected based on the Mg-Al-Ca phase diagram [24,28], to obtain homogeneous microstructures and to avoid the precipitation of further unwanted phases. The forming parameters including the rolling diameter, speed and reduction need to be well controlled to avoid fracture during processing. In the present work, two different thermal mechanical processes were performed to obtain optimal results. Fig. 3 shows the routes of the thermal mechanical processing and the corresponding microstructures of the Mg-2Al-0.2Ca (wt.%) alloy. As shown in Fig. 2(a), the as-cast Mg-2Al-0.2Ca (wt.%) alloy was hot-rolled at 450°C in five passes with 10 % reduction in thickness per pass. Between each pass, the alloy was re-heated at 450 °C for 10 minutes. After the final pass the alloy was annealed at 450 ºC for 15 minutes followed by quenching in water. After hot rolling, a Mg-2Al-0.2Ca (wt.%) alloy (Fig. 2(b)) consisting of columnar grains with an average grain size of 74 ± 6 μm was obtained. However, the EDS measurements (Fig. 2(c)) show that there are intragranular segregation and a small amount of precipitates along the grain boundaries. In order to homogenise the composition and dissolve the precipitates, the Mg-2Al-0.2Ca (wt.%) alloy was subjected to annealing at 500 °C for 24 h after hot-rolling (Fig. 2(d)). After annealing, the precipitates along the grain boundaries and the intragranular segregation are significantly reduced (Figs. 2(e) and 3(f)).

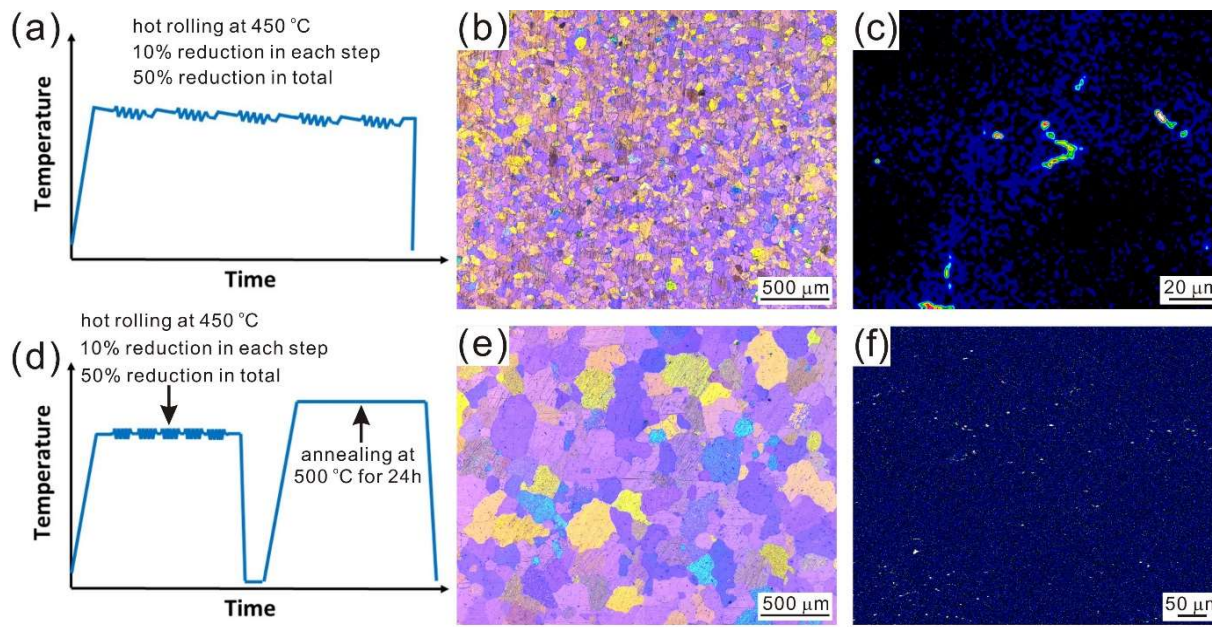

*Fig. 2 (a) Route of thermal mechanical processing of a Mg-2Al-0.2Ca (wt.%) alloy: hot-rolling at 450 ºC in five passes with 10 % reduction in thickness per pass followed by water quenching, and (b) the OM image and (c) EDS map of Al (the areas in red and yellow indicate high concentrations of Al) showing the resulting microstructures. (d) Route of thermal mechanical processing of a Mg-2Al-0.2Ca (wt.%) alloys: hot-rolling at 450 ºC in five passes*

with 10 % reduction in thickness per pass followed by annealing at 500 °C for 24 h, and (e) the OM image and (f) EDS map of Al showing the resulting microstructures.

3.2 Mg-Al-Ca composites

Mg-Al-Ca composites containing a Mg solid solution and a network of the Mg-based Laves phase could also be synthesized by induction melting using a steel crucible under protective Ar atmosphere. It is worth noting that it is inevitable to have Fe contamination in the Mg-Al-Ca alloys due to the solubility of Fe in Al. Since Fe-Al intermetallic layers [30], which hinder the diffusion of Fe into the melt, can be formed between the melt and the steel crucible, Fe contamination is negligible (0.002 – 0.018 wt.% Fe measured by ICP-OES) when the Al content is low (≤6 wt.%). The phases of the Mg-Al-Ca composites can be modified by adjusting the alloy composition [4-7,31]. With increasing Ca/Al ratio from 0 to 1, the type of secondary phases changes from $Mg_{17}Al_{12}$ to the $C15-CaAl_2$, $C36-Ca(Mg,Al)_2$ and $C14-CaMg_2$ Laves phases [5,6]. Moreover, the precipitates become more interconnected and form a skeleton structure [6,7]. In the present work, Mg-Al-Ca composites were melted by induction melting using a steel crucible under Ar atmosphere at 0.8 bar and cast under Ar atmosphere at 0.8 bar in a Cu mould with 10 mm wall thickness and an internal cross section of 25×65 $mm^2$. Representative images of the as-cast Mg-6Al-2Ca, Mg-5Al-3Ca and Mg-4Al-4Ca (wt.%) alloys of composites are shown in Figs. 3(a-c), respectively. The respective Ca/Al ratios of the Mg-6Al-2Ca, Mg-5Al-3Ca and Mg-4Al-4Ca (wt.%) alloys are 0.21, 0.41 and 0.65. It has been confirmed that their main types of intermetallic phases are $C15-CaAl_2$, $C36-Ca(Mg,Al)_2$ and $C14-CaMg_2$, respectively [32]. After heat treatment at 500 °C for 48 h, the interconnected skeletons of Laves phases in the as-cast Mg-Al-Ca composites become rounded and dispersed [32].

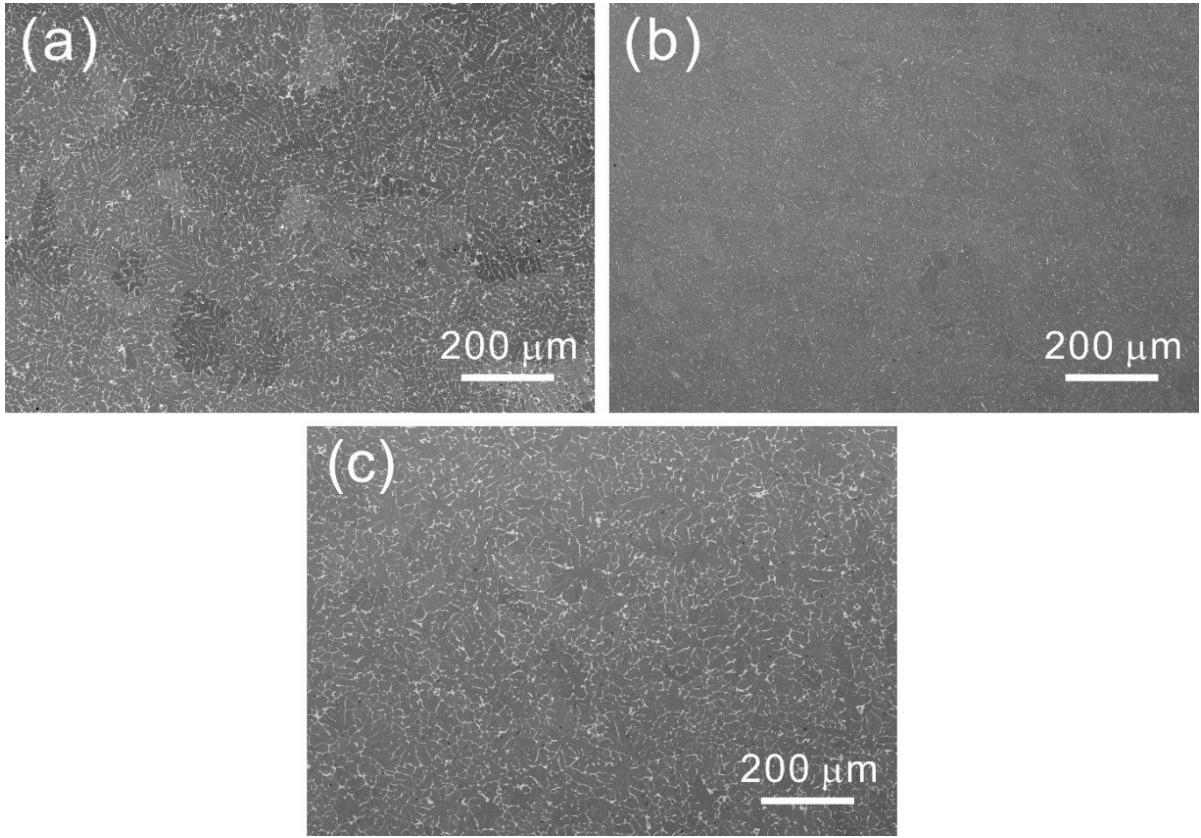

*Fig. 3 Image of the as-cast (a) Mg-6Al-2Ca, (b) Mg-5Al-3Ca and (c) Mg-4Al-4Ca (wt.%) alloys synthesized by induction melting. The intermetallic precipitates in the Mg-6Al-2Ca, Mg-5Al-3Ca and Mg-4Al-4Ca (wt.%) alloys are $C15$-$CaAl_2$, $C36$-$Ca(Mg,Al)_2$ and $C14$-$CaMg_2$ Laves phases, respectively.*

3.3 Mg-Al-Ca intermetallic phases

The phase diagrams of the Mg-Al-Ca ternary system and the Mg-Al, Mg-Ca and Al-Ca binary systems show various types of intermetallic phases. The ternary $C36$-$Ca(Mg,Al)_2$ Laves phase and the binary $C14$-$CaMg_2$, $C15$-$CaAl_2$ and $Mg_{17}Al_{12}$ intermetallic phases are the targeted intermetallic phases in the present work. In order to synthesize the $C14$-$CaMg_2$ Laves phase, the Mg-45Ca (wt.%) alloys were produced from pure Mg and pure Ca by induction melting using a steel crucible. Before melting, the vessel was evacuated and refilled with Ar for several times. In the last step the vessel was evacuated to at least $10^{-5}$ hPa and filled with Ar to 0.8 bar. During melting, the inductive field couples with the ferromagnetic crucible and stirs the melt. The melt was cast into a copper mould with an internal cross section of $30 \times 60$ mm² followed by furnace cooling. The Al-42Ca (wt.%) alloys were prepared from the pure Al and pure Ca by arc-melting on a water-cooled copper hearth in an Ar atmosphere to synthesize the $C15$-$CaAl_2$ Laves phase. After arc-melting, the Al-42Ca (wt.%) alloy was heat treated in a glass tube furnace at 600 °C for 24 h under Ar protection. A Mg-Al diffusion couple was prepared from two blocks of pure Mg and Al to synthesize the $Mg_{17}Al_{12}$ intermetallic phase. The two blocks of pure Mg and Al were grinded up to 400 grit using SiC paper and polished up to 1 μm using diamond paste and ethanol lubricant. They were pressed together and annealed under Ar atmosphere at 400 °C for 1 week, followed by furnace cooling.

The microstructures of the C14-$CaMg_2$ and C15-$CaAl_2$ Laves phases are shown in Figs. 4(a) and 4(b), respectively. While a high density of pores is present in the C14-$CaMg_2$ Laves phase (Fig. 4(a)), the C15-$CaAl_2$ Laves phases (Fig. 4(b)) show homogeneous microstructures and a coarse grain size. The synthesis of $Mg_{17}Al_{12}$ intermetallic phase using the diffusion couple technique was not successful. As shown in Fig. 4(c), neither concentration gradient nor intermetallic phases were obtained in the Mg-Al diffusion couple. The reason for the failure of the Mg-Al diffusion couple might be oxidation of the contact surfaces. Even though the heat treatment was performed in a furnace filled with high purity Ar, oxidation is difficult to avoid during heat treatment in the furnace due to the high reactivity of Mg and Al. Encapsulation of the diffusion couple in a quartz tube might help to reduce oxidation, but to grow thick diffusion layers of the Mg-Al intermetallic phases with coarse grain sizes for dedicated scientific purposes remains challenging.

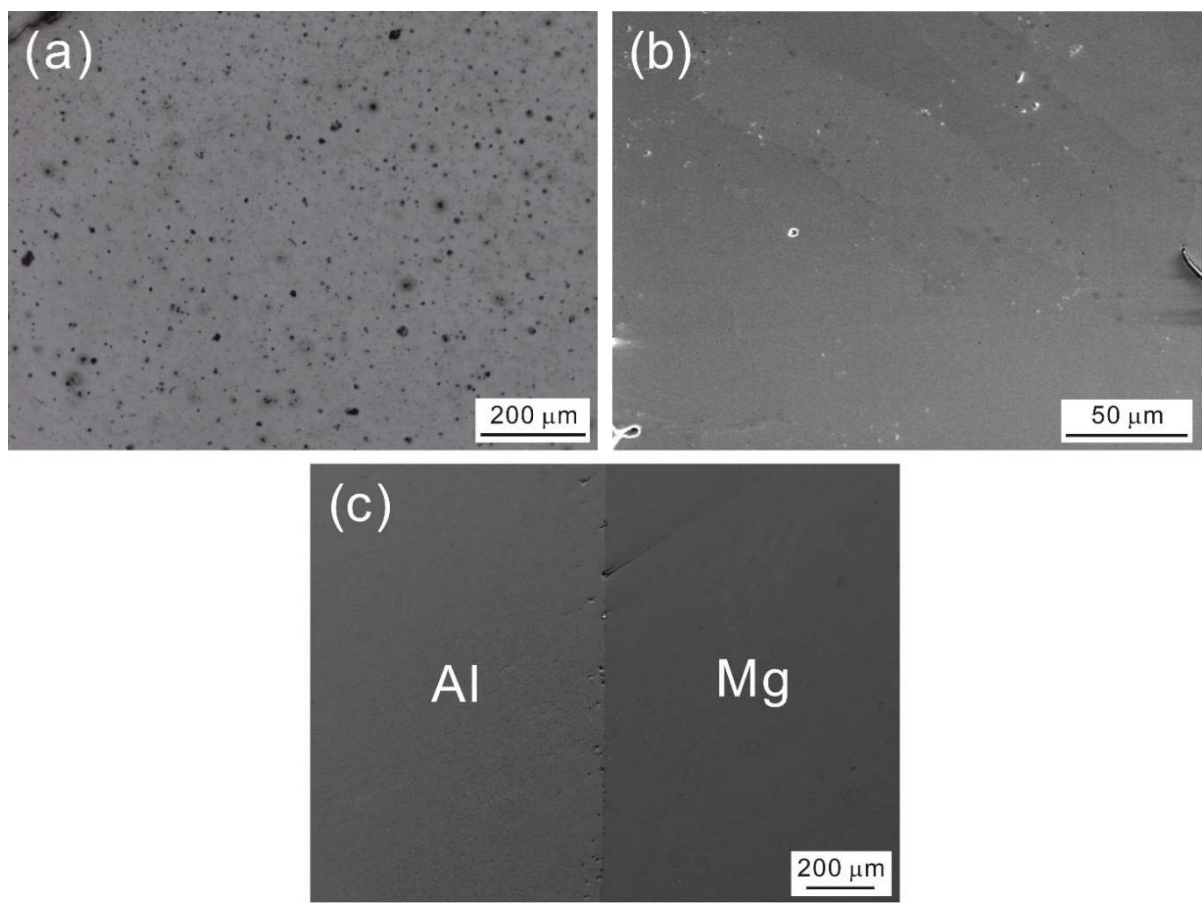

*Fig. 4 SEM images of the (a) C14-CaMg$_2$ Laves phase synthesized by induction melting, (b) C15-CaAl$_2$ Laves phase synthesized by arc-melting, and (c) the Mg-Al diffusion couple after heat treatment at 400 °C for 1 week. No diffusion layer of the Mg$_{17}$Al$_{12}$ intermetallic phases was obtained due to oxidation.*

It has been reported that the Mg-Al-Ca intermetallic phases can be synthesized by combinatorial sputtering [33]. However, desorption of volatile film-forming species during sputtering at elevated temperatures makes the sputtering of C36-Ca(Mg,Al)$_2$ Laves phase challenging. In the present work, different methods were tried to synthesize bulk materials of the C36-Ca(Mg,Al)$_2$ Laves phase. A graphite crucible with boron nitride spray and a sintered boron nitride crucible can be utilized during induction melting to avoid reaction between the melt and the crucible. Besides, a diffusion multiple consisting of pure Mg, Al and Ca and a Mg-CaAl$_2$ liquid-solid diffusion couple were prepared to synthesize the C36-Ca(Mg,Al)$_2$ Laves phase. In addition, the flux growth method [34] and the Bridgman method were also used to synthesize the C36-Ca(Mg,Al)$_2$ Laves phase.

In order to synthesize the C36-Ca(Mg,Al)$_2$ Laves phase, Mg-30Al-44Ca (wt.%) alloys were prepared by induction melting using a graphite crucible with boron nitride spray under Ar protection. The graphite crucible was dissolved during melting, and thus it was replaced by a sintered boron nitride crucible. However, the sintered boron nitride crucible is so brittle that it fractured during synthesis. Since no suitable crucible can be used, manual induction melting in a quartz tube was performed to synthesize the C36-Ca(Mg,Al)$_2$ Laves phase instead. Pure Mg, Al, Ca were mixed and placed in a quartz tube. The quartz tube was evacuated and back-filled with Ar. The alloys were re-melted three times to ensure homogeneity. In addition, a Mg-Al-Ca diffusion multiple was prepared from two blocks of pure Mg and Al with a thin Ca foil in between. The Ca foil was produced by melting and cold rolling. However, due to the oxidation of Ca the diffusion multiple failed. A Mg-CaAl$_2$ liquid-solid was prepared to synthesize the C36-Ca(Mg,Al)$_2$ Laves phase instead. A block of Mg was placed on top of a C15-CaAl$_2$ Laves phase alloy and melt inside an arc-melting furnace under protective Ar atmosphere. While Mg melted first due to the lower melting point, the C15-CaAl$_2$ Laves phase alloy remained solid and therefore formed a liquid-solid diffusion couple with the molten Mg. The composition of the Mg-30Al-44Ca (wt.%) alloy prepared by manual induction melting significantly deviates from the nominal composition, presumably due to a Mg loss during induction melting. As shown in Fig. 5(a), the Mg-30Al-44Ca (wt.%) alloy mainly contains Mg and C15-CaAl$_2$ Laves phase. Fig. 5(b) shows that the diffusion zone of the Mg-CaAl$_2$ liquid-solid diffusion couple mainly contains dendrites of C15-CaAl$_2$ and a eutectic mixture of Mg and C15-CaAl$_2$. Thus, the synthesis of the C36-Ca(Mg,Al)$_2$ Laves phase by manual induction melting and liquid-solid diffusion couple methods was not successful.

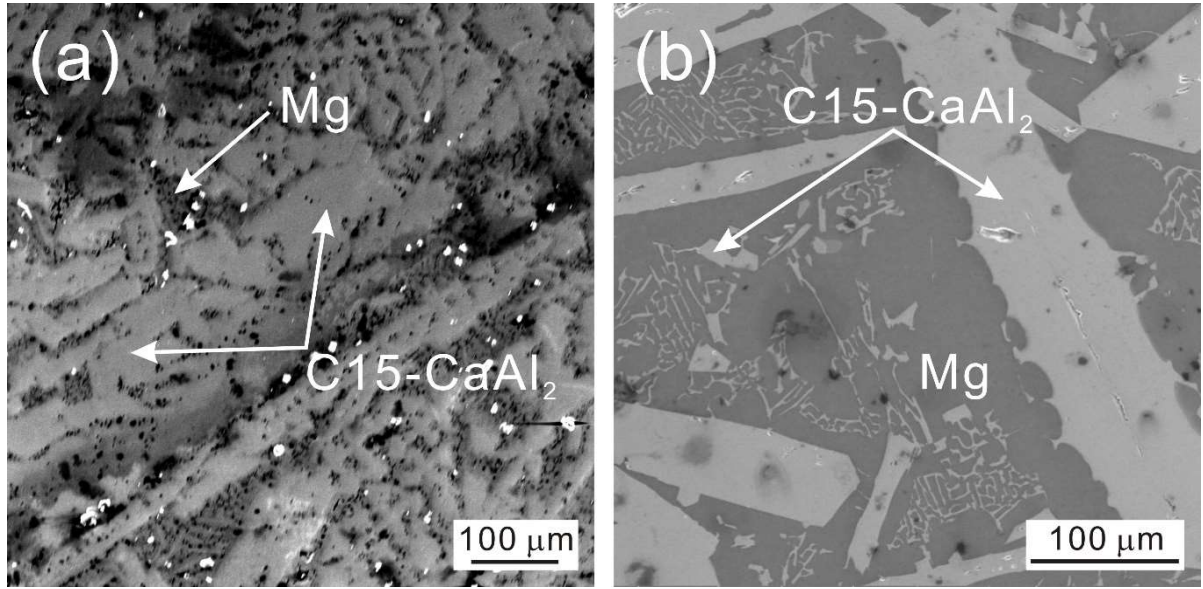

*Fig. 5 SEM images of the (a) Mg-30Al-44Ca (wt.%) alloy synthesized by manual induction melting, and (b) Mg-CaAl$_2$ liquid-solid diffusion couple.*

The flux growth method [34] was then used to synthesize the C36-Ca(Mg,Al)$_2$ Laves phase. Figs. 6(a) and 6(b) shows a schematic illustration of the setup. The starting materials of a Mg-30Al-44Ca (wt.%) alloy were mixed and placed in an alumina crucible with a mesh fixed at a position above the alloy for decanting the melt at the end of crystal growth. The entire crucible was encapsulated using a Quartz tube filled with Ar (Fig. 6(c)). The quartz tube was heated up to 870 ºC and annealed for 2 h to melt the starting materials. After melting, the quartz tube was cooled down to 850 ºC at 10 ºC/h and then slowly cooled down to 750 ºC at 1 ºC/h, followed by centrifugation to spin off the remaining liquid phase. According to the phase diagram [24,28,35], at the composition of Mg-30Al-44Ca (wt.%) the C36-Ca(Mg,Al)$_2$ Laves phase is in equilibrium with the liquid phase at 750 ºC. Therefore, after removing liquid phase, the crystal of the C36-Ca(Mg,Al)$_2$ Laves phase can be obtained. However, as shown in Fig. 6(d), after cooling the quartz tube was blackened on the inside and showed radial cracking at the position where the crucibles met. After cracking the tube we found one ingot (Fig. 6(e)) in the crucible and no materials on the grid. It suggests that the whole melt solidified above 750°C and there was no liquid left to be centrifuged into the upper crucible. The fact that there was no liquid phase present at 750°C could either be due to an error in the phase diagram, a significant change in melt composition due to evaporation losses, or a temperature offset between our thermocouple and the actual temperature in the crucible.

Finally, we tried to synthesize the C36-Ca(Mg,Al)$_2$ Laves phase by the Bridgman method. Pure elements at composition Mg-28Al-46Ca (wt.%) were placed in a cylindrical tantalum crucible of 14 mm in diameter and 70 mm in length. The crucible was sealed with 0.6 bar Argon and placed on a water-cooled cold finger in a Bridgman apparatus, the chamber of which was then heated to 900 °C. After temperature equilibration the actual growth was carried out by lowering the crucible out of the hot zone at a velocity of 5 mm/h.

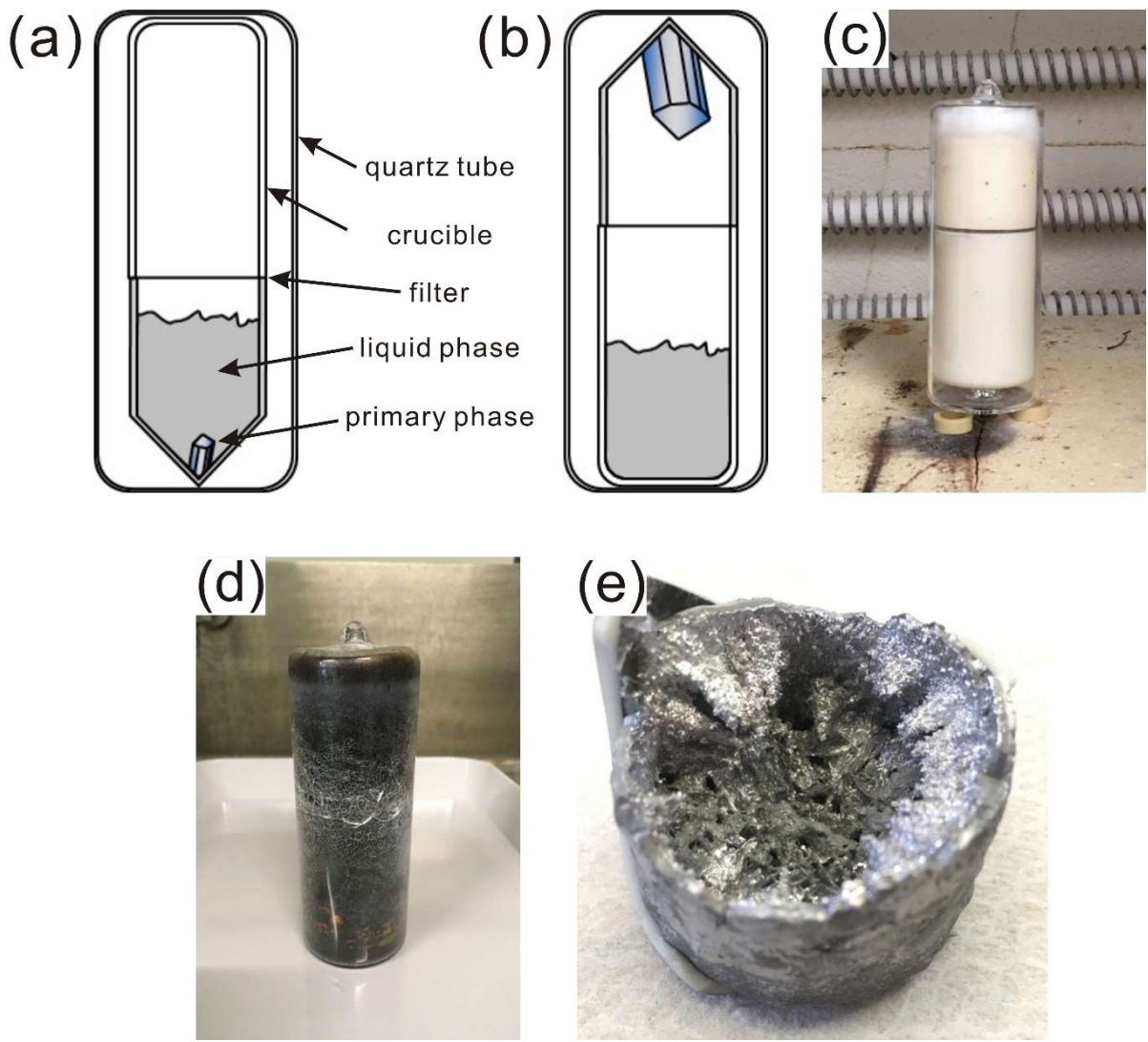

*Fig. 6 (a) A schematic illustration of the set-up of the solution growth method. (b) After heat treatment, the primary phase was separated from the remaining liquid phase by centrifugation. (c) An alumina crucible encapsulated in a quartz tube filled with Ar. (d) After cooling the quartz tube was blackened on the inside and showed radial cracks. (e) Only one ingot was obtained in the crucible and no materials were left on the grid.*

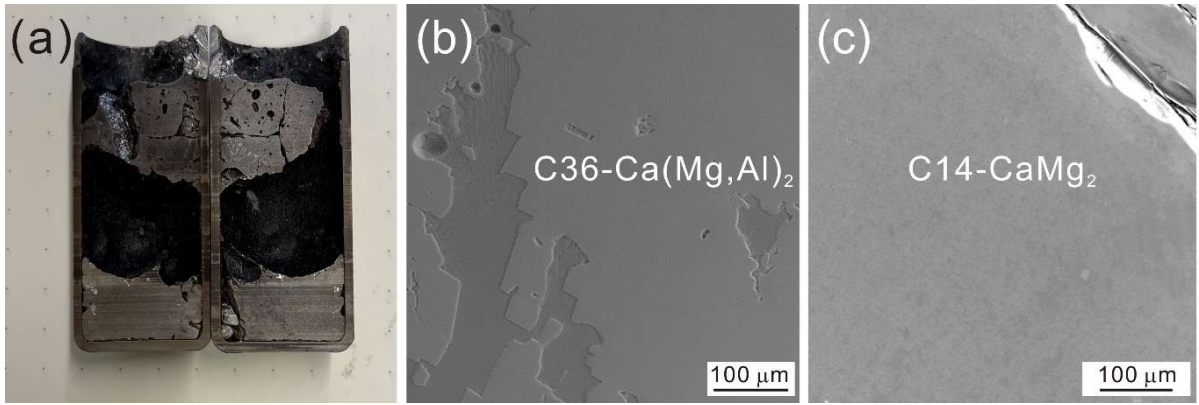

Fig. 7 (a) The Mg-28Al-46Ca (wt.%) alloy synthesized by the Bridgman method to obtain the C36-Ca(Mg,Al)$_2$ Laves phase. SEM images of the (b) top and (c) bottom parts of the sample.

As shown in Fig. 7(a), the Mg-28Al-46Ca (wt.%) alloy synthesized by the Bridgman method contains two parts and a large void in the middle. The reason why and how the large void occurs are not clear yet. The microstructures of the top and the bottom parts of the sample obtained by the Bridgman method are shown in Figs. 7(b) and 7(c), respectively. Their average compositions measured by EDS are Mg-45.37Al-38.63Ca (at.%) and Mg-2.27Al-39.40Ca (at.%), respectively. The EBSD measurements indicate that the top and the bottom parts of the sample are the C36-Ca(Mg,Al)$_2$ and C14-CaMg$_2$, respectively. Therefore, it is possible to obtain the C36-Ca(Mg,Al)$_2$ Laves phase after a time-consuming process of trial and error but the synthesis of flawless single-phase bulk materials of the C36-Ca(Mg,Al)$_2$ Laves phase is very difficult.

## 4 Conclusions

Due to the high reactivity and high vapor pressures of Mg and Ca and high affinity of Al with Fe, the synthesis of bulk materials of Mg-Al-Ca solid solutions, composites and intermetallic phases for specific scientific purposes is rather challenging. Different methods were investigated in the present work to show the challenges and pitfalls in the synthesis of the Mg-Al-Ca bulk materials. The following conclusions are drawn:

(1) To reduce the evaporation of Mg and Ca, the Mg-Al-Ca solid solutions can be synthesized by induction melting using a steel crucible under Ar atmosphere with high pressure. During melting and casting, the high pressure Ar atmosphere can reduce casting defects and improve mechanical properties of the as-cast Mg-Al-Ca alloys. The precipitation along grain boundaries and the intragranular segregation in the Mg-Al-Ca solid solutions can be significantly reduced by hot rolling and homogenization annealing.

(2) The Mg-Al-Ca composites comprising a Mg matrix and a network of the Mg-based Laves phase can be synthesized by induction melting under Ar atmosphere. When the Al content is low (≤6 wt.%), the Fe contamination in the Mg-Al-Ca composites is negligible, and thus steel crucible can still be used during melting. The phases and microstructures of the Mg-Al-Ca composites can be adjusted the alloy composition and heat treatment.

(3) While the C14-CaMg$_2$ Laves phase can be synthesized by induction melting using a steel crucible under Ar protection, the C15-CaAl$_2$ Laves phase can be prepared by arc melting. However, the synthesis of Mg$_{17}$Al$_{12}$ intermetallic phase by the diffusion couple technique was not successful due to the oxidation of the contact surfaces during heat treatment. The synthesis of the C36-Ca(Mg,Al)$_2$ Laves phase by manual induction melting failed, due to a high loss of Mg during melting. No C36-Ca(Mg,Al)$_2$ Laves phase can be obtained from the liquid-solid Mg-CaAl$_2$ diffusion couples, either. It is possible to synthesize the C36-Ca(Mg,Al)$_2$ Laves phase by the Bridgman method but it is challenging to produce flawless single-phase bulk materials of the C36-Ca(Mg,Al)$_2$ Laves phase.

**Acknowledgements:**

We thank Mr. Jürgen Wichert and Mr. Michael Kulse for their technical supports in alloy synthesis. This work was supported by the German research foundation (DFG) within the Collaborative Research Centre SFB 1394 "Structural and Chemical Atomic Complexity—From Defect Phase Diagrams to Materials Properties" (Project ID 409476157). H. Springer wishes to acknowledge funding through the Heisenberg-program of the Deutsche Forschungsgemeinschaft (Project ID 416498847).

**Author contributions:**

L. Tanure and H. Springer conceived and designed the experiments. W. Luo wrote the manuscript. M. Felten, D. Zander and W. Delis analysed the microstructures of Mg-Al-Ca solid solutions. N. Aye performed characterizations of the Mg-Al-Ca composites. M. Freund contributed to the characterizations of the Mg-Al-Ca intermetallics. C. Thomas and M. Feuerbacher contributed to the synthesis of the C36-Ca(Mg,Al)$_2$ Laves phase. S. Sandlöbes-Haut and S. Korte-Kerzel provided feedback and suggestions to improve the experiments. All authors contributed to discussion of the results and reviewed the manuscript.

**Conflicts of Interest:** The authors declare no conflict of interest.